\documentclass[aps,prl,reprint,superscriptaddress,amsmath,amssymb,floatfix,showpacs]{revtex4-1}
\usepackage{times}
\usepackage{graphicx}
\usepackage{subfigure}
\usepackage{color}
\usepackage{epstopdf}
\usepackage{slashed}
\usepackage{bm}
\usepackage{mathtools}
\usepackage{braket}
\usepackage{soul}

\begin{document}

\title{Antiferroquadrupolar and Ising-Nematic Orders of a Frustrated Bilinear-Biquadratic
Heisenberg Model and Implications for the Magnetism of FeSe}

\author{Rong Yu}
\affiliation{Department of Physics and Beijing Key Laboratory of Opto-electronic Functional Materials \& Micro-nano Devices, Renmin University of China, Beijing 100872, China
}
\affiliation{Department of Physics and Astronomy, Collaborative Innovation
Center of Advanced Microstructures, Shanghai Jiaotong University, Shanghai
200240, China}

\author{Qimiao Si}
\affiliation{Department of Physics \& Astronomy, Rice University, Houston, Texas 77005}

\begin{abstract}
Motivated by the properties of the iron chalcogenides, we study the phase diagram of a
generalized Heisenberg model with frustrated bilinear-biquadratic interactions on a square lattice.
We identify zero-temperature phases with antiferroquadrupolar and Ising-nematic orders.
The effects of quantum fluctuations  and interlayer couplings are analyzed.
We propose the Ising-nematic order as underlying the structural phase
transition observed in the normal state of FeSe, and discuss the
role of the Goldstone modes of the antiferroquadrupolar order
for the dipolar magnetic fluctuations in this system.
Our results provide a considerably broadened perspective on the overall
magnetic phase diagram of the iron chalcogenides and pnictides,
and are amenable to tests by new experiments.
\end{abstract}


\maketitle

{\it Introduction.~}
Because superconductivity develops near magnetic order in most of
the iron pnictides and chalcogenides, it is important to
understand the nature of their magnetism. The iron pnictide
families typically have parent compounds that show a collinear $(\pi,0)$
antiferromagnetic order~\cite{PCDai_review12}. Lowering the temperature
in the parent compounds gives rise to
a tetragonal-to-orthorhombic distortion, and the
temperature $T_s$ for this structural transition
is either equal to or larger than the
N\'{e}el transition temperature $T_N$. A likely explanation for $T_s$ is
an Ising-nematic transition at the electronic level. It was
recognized from the beginning that models with quasi-local moments
and their frustrated Heisenberg $J_1-J_2$ interactions
\cite{SiAbrahams08} feature such an Ising-nematic transition
\cite{Chandra89,Fang08,Xu08,JDai09}.
Similar conclusions have subsequently been reached
in models that are based on Fermi-surface instabilities
\cite{Fernandes14}.

The magnetic origin for the nematicity fits well with the
experimental observations of the spin excitation spectrum observed
in the iron pnictides. Inelastic neutron scattering experiments
\cite{Diallo10} in the parent iron pnictides have revealed a low-energy
spin spectrum whose equal-intensity contours in the wavevector space
form ellipses near $(\pm \pi,0)$ and $(0, \pm \pi)$. At high energies,
spin-wave-like excitations are observed all the way to the boundaries of the
underlying antiferromagnetic Brillouin zone \cite{Harriger11}. These
features are well captured by Heisenberg models with
the frustrated $J_1-J_2$ interactions and biquadratic couplings
\cite{Yu2012,Wysochi11}, although at a refined level it is also important
to incorporate the damping provided by the coherent itinerant fermions
near the Fermi energy \cite{Yu2012}.

Experiments in bulk FeSe do not seem to fit into this framework.
FeSe is one of the canonical iron chalcogenide
superconductors~\cite{Wu08,Mao08}. In the single-layer limit,
it currently holds particular promise towards a very
high $T_c$ superconductivity~\cite{Xue12,ZXShen14,Jia14} driven
by strong correlations~\cite{XJZhou-pnas14}.
In the bulk form, this compound displays a tetragonal-to-orthorhombic structural
transition, with $T_s \approx 90$ K, but no N\'{e}el transition has been
detected \cite{McQueen09,Medvedev09,Bohmer14,Baek14}.
This distinction has been interpreted as pointing
towards the failure of the magnetism-based origin for the
structural phase transition~\cite{Bohmer14,Baek14}.
At the same time, experiments have also revealed another aspect of the emerging puzzle.
The structural transition clearly induces
dipolar magnetic fluctuations \cite{Bohmer14,Baek14}.

In this Letter, we show that a generalized Heisenberg model with frustrated bilinear-biquadratic
couplings on a square lattice contains a phase
with both a $(\pi,0)$ antiferroquadrupolar (AFQ) order and
an Ising-nematic order.
The model in this regime displays a finite-temperature transition into the Ising-nematic order and,
in the presence of interlayer couplings,
also a finite-temperature transition into the AFQ order.
We suggest that such physics is compatible with the aforementioned and related properties of FeSe.
The Goldstone modes of the AFQ order
 are responsible for the onset of
 dipolar magnetic fluctuations
near the wave-vector $(\pi,0)$, which is experimentally testable.

{\it Model.~} We consider a
spin
Hamiltonian with $S\geqslant1$ on a two-dimensional (2D) square lattice: \begin{align}\label{Eq:HamTot}
 H &= \frac{1}{2}\sum_{i,\delta_n}
 \left\{J_n \mathbf{S}_i\cdot\mathbf{S}_j + K_n (\mathbf{S}_i\cdot\mathbf{S}_j)^2\right\},
\end{align}
where $j=i+\delta_n$, and $\delta_n$ connects site $i$ and its $n$-th
nearest neighbor sites with $n=1$, $2$, $3$. Here $J_n$ and $K_n$
are, respectively, the bilinear and biquadratic couplings between the $n$-th
nearest neighbor spins.
For iron pnictides and iron chalcogenides, the local moments describe the spin degrees of freedom
associated with the incoherent part of the electronic excitations
and reflect the bad-metal behavior of these systems\cite{PCDai_review12,SiAbrahams08,Fang08,Xu08,JDai09}.
A nonzero $J_3$ is believed to be important for the iron chalcogenides, especially
FeTe \cite{Ma_etal_09}.
The biquadratic couplings $K_n$ are expected to play a significant role in multi-orbital systems
with moderate Hund's coupling \cite{Fazekas}.
The nearest neighbor
coupling $K_1$ was included in previous
studies ~\cite{Wysochi11,Yu2012} to understand the strong anisotropic spin excitations
in the Ising-nematic ordered phase, where the ground state has a $(\pi,0)/(0,\pi)$
antiferromagnetic (AFM) order.
With the goal of searching for the new physics that could describe the properties of FeSe,
in this work, we take these couplings as variables and study the pertinent unusual phases
in the phase diagram.
In the following, to simplify the discussion on the relevant AFM and AFQ phases,
we take $K_1=-1$ and use $|K_1|$ as the energy unit.
Note that a moderately positive $K_1$ in the presence of further-neighbor $K_n$ couplings will
lead to similar results, but a $K_1$ coupling alone in the absence of the latter
will not generate the physics discussed below.

Some general considerations are in order.
For $S\geqslant1$
\begin{equation}\label{Eq:QdotQ}
(\mathbf{S}_i\cdot\mathbf{S}_j)^2 = \frac{1}{2} \mathbf{Q}_i \cdot \mathbf{Q}_j
- \frac{1}{2} \mathbf{S}_i\cdot\mathbf{S}_j
+\frac{1}{3} \mathbf{S}_i^2\mathbf{S}_j^2,
\end{equation}
where $\mathbf{Q}_i$ is a quadrupolar operator with five components
$Q_i^{r^2-3z^2}=\frac{1}{\sqrt{3}} [(S_i^x)^2+(S_i^y)^2-2(S_i^z)^2]$,
$Q_i^{x^2-y^2}=(S_i^x)^2-(S_i^y)^2$, $Q_i^{xy}=S_i^x S_i^y + S_i^y S_i^x$,
$Q_i^{yz}=S_i^y S_i^z + S_i^z S_i^y$, and $Q_i^{zx}
=S_i^z S_i^x + S_i^x S_i^z$.
Therefore, the biquadratic interaction may promote a long-range
ferroquadrupolar-antiferroquadrupolar (FQ-AFQ) order.
With the aforementioned motivation, we are interested in
a $(\pi,0)$ AFQ order,
which
would break the C$_4$ symmetry and
should yield
an Ising-nematic order parameter.

\begin{figure}[t!]
\includegraphics[width=85mm]{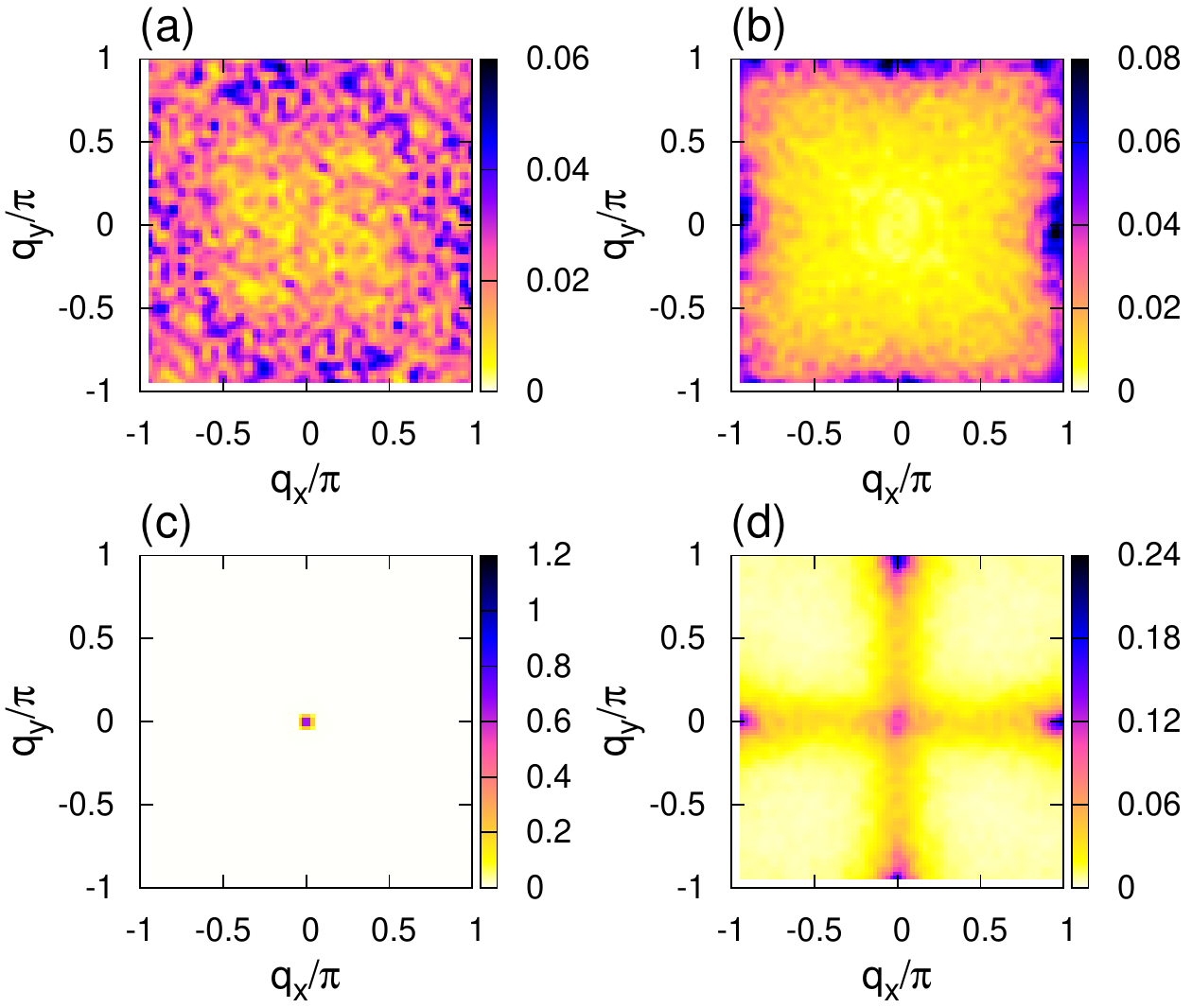}
\caption{Momentum distribution of the dipolar (top row) and quadrupolar (bottom row) magnetic
structure factors at $K_2=-1$ (in (a) and (c)) and $K_2=1.5$ (in (b) and (d)), respectively.
Here, $J_1=J_2=1$, $J_3=K_3=0$, and $K_1=-1$. The calculations are done
on a 40$\times$40 lattice at $T/|K_1|$=0.01 with up to $10^5$ Monte Carlo steps.
In (d), besides the leading
AFQ
correlations at $(\pi,0)$ and $(0,\pi)$, subleading
FQ
correlations are observed at finite temperatures;
as the temperature is lowered,
the former
is enhanced whereas
the latter is
is diminished.
}
\label{Fig:SSF}
\end{figure}

{\it Low-temperature phase diagram of the classical spin model.~}
We first study the model in Eq.~\eqref{Eq:HamTot}
for classical spins. For simplicity, we discuss the case $K_3=0$.
We have calculated the dipolar and quadrupolar magnetic structure factors via Monte Carlo simulations
using the standard Metropolis algorithm.\cite{MCBook}
Representative results for the structure factor data are shown in Fig.~\ref{Fig:SSF}, for
$J_3=0$ and $J_1=J_2$.
The two cases, corresponding to different values of $K_2$, show, respectively, dominant
ferroquadrupolar (FQ) and $(\pi,0)$ AFQ correlations,
for the finite-size systems studied and at a very low temperature $T/|K_1|=0.01$.

\begin{figure}[t!]
\includegraphics[width=65mm]{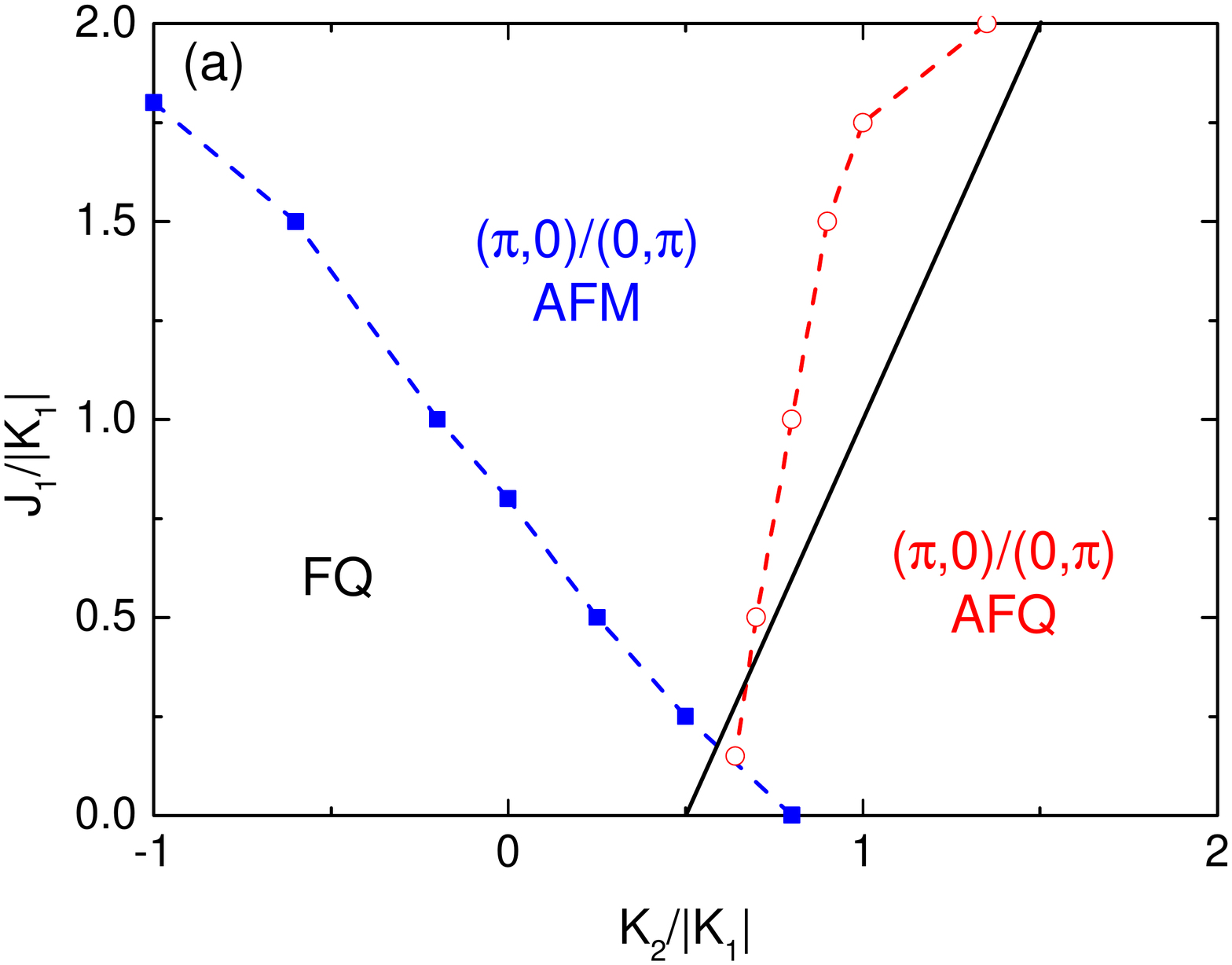}
\includegraphics[width=65mm]{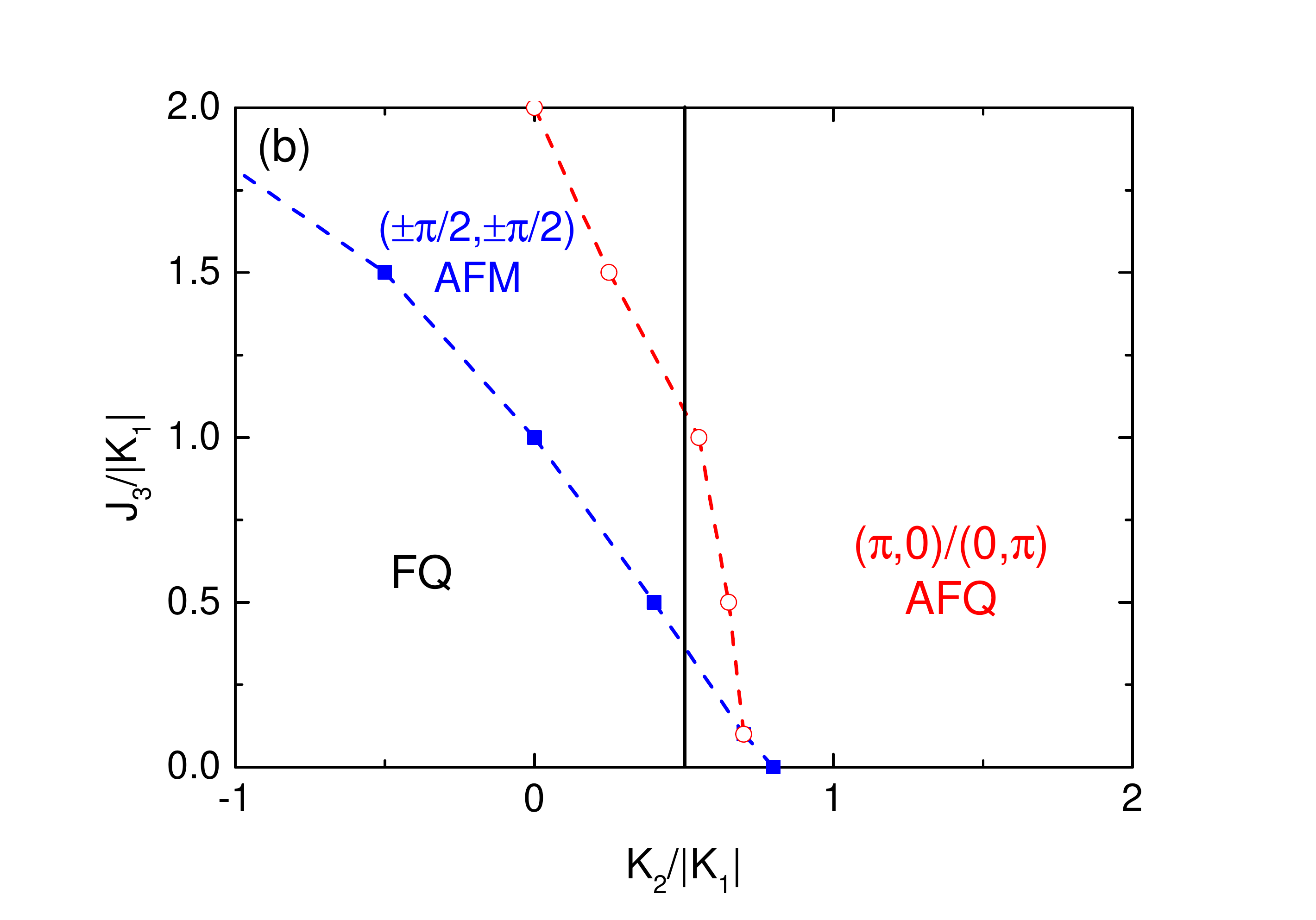}
\caption{Low-temperature phase diagrams of the classical 
bilinear-biquadratic Heisenberg
model at (a): $J_1=J_2$, $J_3=K_3=0$ and (b): $J_1=K_3=0$, $J_2=J_3$.
Both are shown at $T/|K_1|=0.01$. Dashed lines show finite-temperature crossovers between different orders.
The dominant order in each regime is labeled.
In each case, the solid line shows the mean-field phase boundary at $T=0$. }
\label{Fig:Crossover}
\end{figure}

Overall, as shown in Fig.~\ref{Fig:Crossover}(a), we find that there are large regimes in the phase diagram
in which the FQ and $(\pi,0)$ AFQ moments are almost ordered, while the dipolar moments
coexisting with the FQ/AFQ moments
are very weakly correlated. Hence in these regimes, the dominant low-temperature order is the FQ/AFQ one.
In between these, there is a regime in which the dominant correlation occurs in the $(\pi,0)$
AFM
channel.

Similar results for the case of $J_1=0$ and $J_2=J_3$
are shown in Fig.~\ref{Fig:Crossover}(b).
A large regime with dominating
FQ or $(\pi,0)$ AFQ correlations is also found. The difference from the case
of $J_3=0$ and $J_1=J_2$ occurs in the regime
with dominant AFM correlations, for which the wavevector is now $(\pm\pi/2,\pm\pi/2)$
as relevant to the FeTe compound.

For 2D systems,
thermal fluctuations will ultimately (in the thermodynamic limit) destroy
any order that breaks a continuous global symmetry
at any nonzero temperature~\cite{MERMIN-WAGNER}.
The dashed lines in Fig.~\ref{Fig:Crossover} therefore mark crossovers between regimes
with different dominant correlations.
At $T=0$, on the other hand, genuine FQ/AFQ can occur in our model on the square lattice.
We have therefore also analyzed the mean-field phase diagrams at $T=0$. The resulting phase boundary
is shown in each case as a solid line
in Fig.~\ref{Fig:Crossover}. The results are compatible with the crossovers identified
at low  but nonzero temperatures.
For the case of $J_3=0$ and $J_1=J_2$, shown in Fig.~\ref{Fig:Crossover}(a), the phase on the left
of the solid line
has a mixture of
an AFM phase ordered at $\mathbf{q}=(\pi,0)/(0,\pi)$
and a FQ phase.
The phase on the right of the solid line has an AFQ phase ordered at
$\mathbf{q}=(\pi,0)/(0,\pi)$.
Note that in the classical limit, the spins are treated as O(3) vectors, and should always
be ordered at zero temperature.
We find that in the AFQ phase, the spins can be ordered at
a wavevector $(q,\pi)/(\pi,q)$ for arbitrary $q$, with an infinite degeneracy.\cite{SM}
Such a frustration would likely stabilize a purely AFQ ground state when quantum fluctuations
are taken into account (see below).
For the case of $J_1=0$ and $J_2=J_3$, shown in Fig.~\ref{Fig:Crossover}(b),
the mean-field result also yields FQ or $(\pi,0)$ AFQ,
respectively, to the left or right of the solid line.
However, the wave vector for the AFM orders that mix, respectively, with the FQ and $(\pi,0)$ AFQ order
has become $(\pm \pi/2,\pm \pi/2)$.\cite{SM}

\begin{figure}[t!]
\includegraphics[width=80mm]{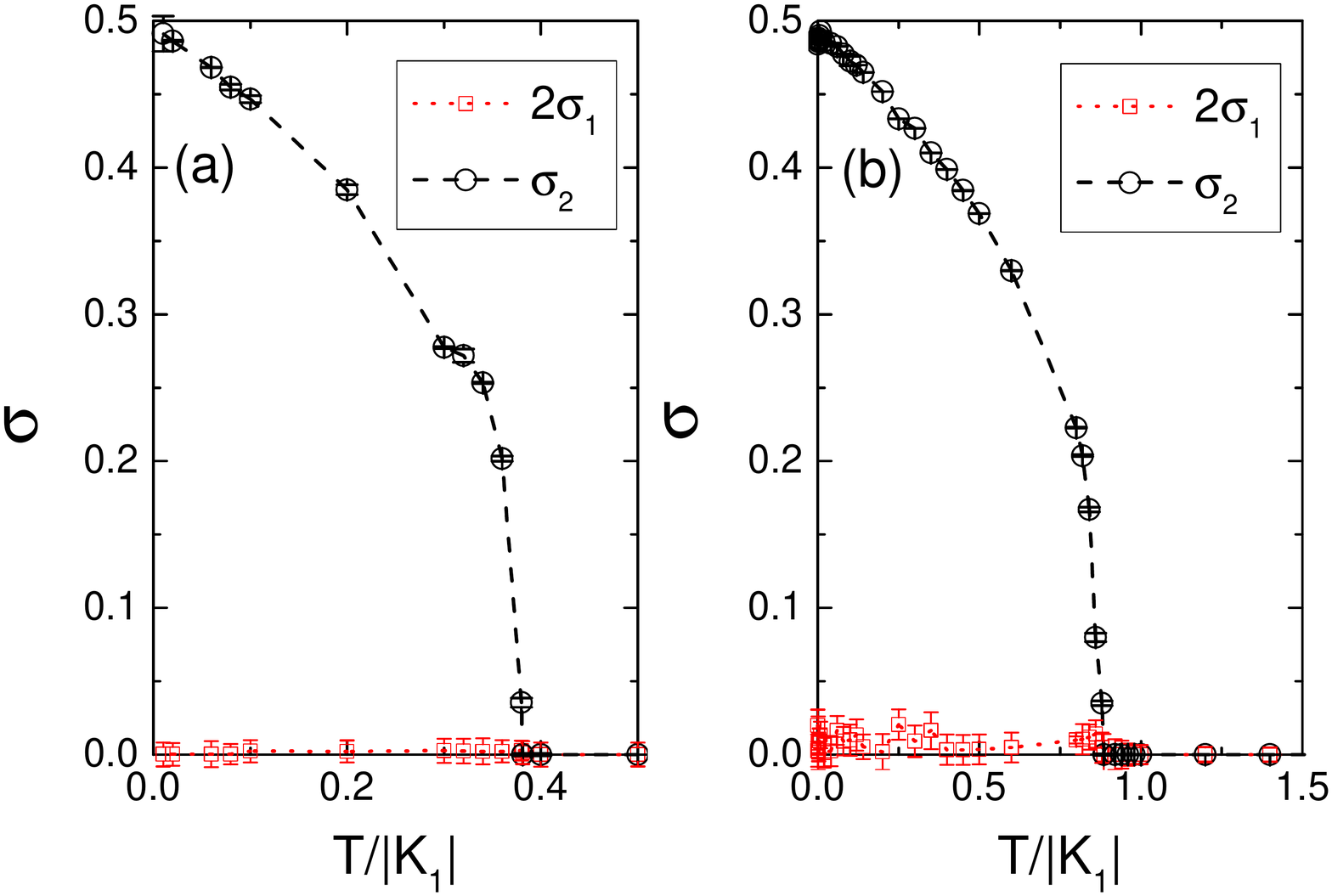}
\caption{Temperature dependence of the Ising-nematic order parameters $\sigma_1$ and $\sigma_2$
at (a) $J_1=J_2=J_3=0$, $K_1=-1$, and $K_2=1$ and (b) $J_1=0$, $J_2=J_3=0.5$, $K_1=-1$,
and $K_2=2$. In both cases the dominant part of the Ising-nematic order is $\sigma_2$,
which is associated with the AFQ order.}
\label{Fig:Ising2}
\end{figure}

Similar to the $(\pi,0)$
AFM state, the $(\pi,0)$
AFQ phase breaks the lattice $C_4$ rotational symmetry.
An
accompanying
Ising-nematic transition is
to be
expected,
and should develop
at nonzero temperatures even in two dimensions.
We define the general Ising-nematic operators as follows:
\begin{equation}
 \sigma_n = (\mathbf{S}_i\cdot\mathbf{S}_{i+\hat{x}})^n - (\mathbf{S}_i\cdot\mathbf{S}_{i+\hat{y}})^n,
\end{equation}
where $n=1,2$. We also introduce the quadrupolar $\mathbf{Q}_{A/B}$
to be the linear superposition of
$\mathbf{Q}(\pi,0)/(0,\pi)$, with the ratios of their coefficients to be $\pm 1$ respectively.
From Eq.~\eqref{Eq:QdotQ}, we see that for quantum spins,
the Ising-nematic order associated with $\mathbf{Q}$ should be seen in both
$\sigma_1$ and $\sigma_2$.
For classical spins, since $\mathbf{Q}_i \cdot \mathbf{Q}_j = 2(\mathbf{S}_i\cdot\mathbf{S}_j)^2
- \frac{2}{3} \mathbf{S}_i^2\mathbf{S}_j^2$,
only $\sigma_2$
will manifest
$\mathbf{Q}_A\cdot\mathbf{Q}_B$.
This allows us to determine the origin of the Ising-nematic order in the AFQ+AFM phase.
As shown in Fig.~\ref{Fig:Ising2}(a),
for the $K_1-K_2$ model,
$\sigma_2$ is ordered at $T_{\sigma}/|K_1|\approx0.38$
but $\sigma_1\approx0$ for any $T>0$.
Likewise from Fig.~\ref{Fig:Ising2}(b), in the case $J_1=0$ and $J_2=J_3$,
the dominant Ising nematic order parameter is $\sigma_2$
for $T<T_{\sigma}\approx 0.9$, and $\sigma_1$
never
becomes substantial
down to the lowest
temperature $T=10^{-4}$
accessible to our numerical simulation. These indicate that the Ising-nematic order
in the AFQ+AFM phase is associated with the anisotropic spin quadrupolar fluctuations.

\begin{figure}[t!]
\includegraphics[width=80mm]{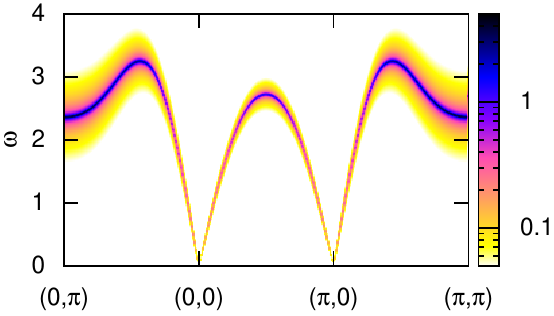}
\caption{Calculated spin excitation spectrum in the $(\pi,0)$ AFQ phase of the quantum $S=1$ model.
We have taken $J_1=J_2=0.25$, $J_3=0$, $K_2=0.5$, and $K_3=-0.3$.
The color codes the dynamical spin dipolar structure factor, $\sqrt{S^{xx}_D(\mathbf{q},\omega)}$.
}
\label{Fig:SpinExcitation}
\end{figure}

{\it The quantum spin models.~}
The AFQ phase and the associated
Ising-nematic transition are features of the generalized $J-K$ model for both
classical and quantum spins.
To consider the effect of quantum fluctuations, we consider
the case of $S=1$.
We study
its ground-state
properties via a semiclassical variational approach by using
an
SU(3)
representation~\cite{LauchliMilaPenc06},
and identify
parameter regimes
that stabilize
the AFQ phase.
We further study the spin excitations in the AFQ phase with the ordering wavevector $\mathbf{q}_A=(\pi,0)$
using a flavor-wave theory.\cite{SM}
Because the
AFQ order breaks the continuous spin-rotational invariance,
the
Goldstone modes
will
have a nonzero dipolar matrix element
\cite{LauchliMilaPenc06,Tsunetsugu06}.
To explicitly demonstrate this, we calculate the
dynamical spin dipolar structure factor $
S^{xx}_D(\mathbf{q},\omega)$ near $\mathbf{q}_A$,
which is shown in Fig.~\ref{Fig:SpinExcitation}.
Therefore, the development of the AFQ order is accompanied by a sharp rise
in the dynamical spin dipolar correlations
centered around the wavevector $(\pi,0)$ (and symmetry-related wave vectors).

{\it Coupling to itinerant fermions and interaction between layers.~}
One additional effect of the quantum fluctuations is that it can suppress the weak
AFM order when the dominant order is AFQ. We discuss one source of such an effect,
which is the coupling of the order parameters to the coherent itinerant fermions.
The effect of coupling to the itinerant fermions can be treated as in Ref.~\cite{JDai09}
within
an effective Ginzburg-Landau action, and is briefly discussed
 in the Supplemental Material \cite{SM}.
 When only the $(\pi,0)$ AFM order and the Ising-nematic order are present,
 the coupling to the itinerant fermions will suppress the AFM and Ising-nematic order
 concurrently \cite{JWu14}.
However, when the dominant order is AFQ,
the coupling to the itinerant fermions can suppress the AFM order
while retaining the stronger AFQ order and the associated Ising-nematic order.

\begin{figure}[t!]
\includegraphics[width=80mm]{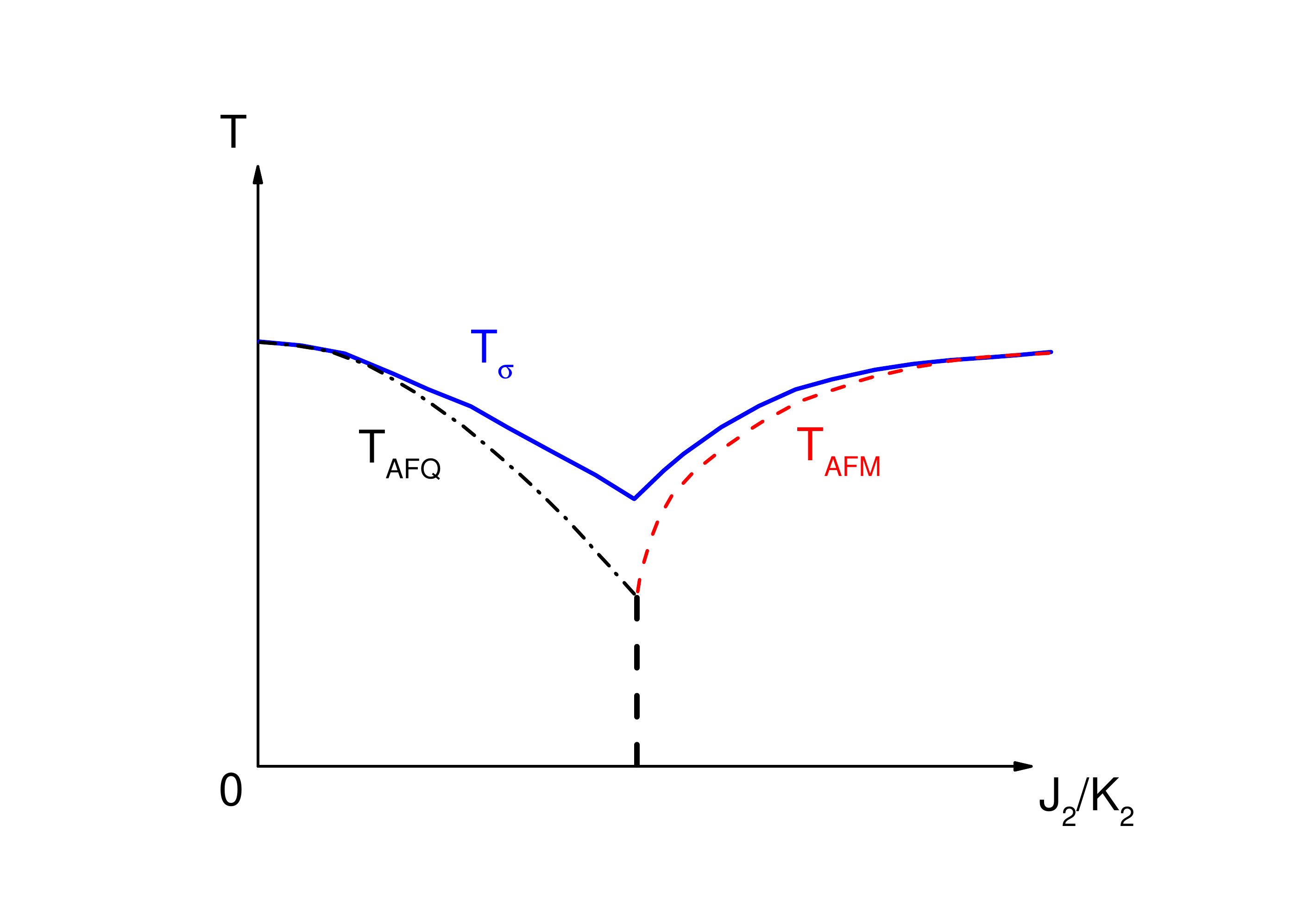}
\caption{Skecthed phase diagram in terms of $T$ and $J_2/K_2$.
The dominant order may be either AFQ or  AFM. The thinner dashed lines show
the associated ordering temperautures $T_{AFQ}$ and $T_{AFM}$. A first-order transition
(thicker dashed line) takes place at an intermediate $J_2/K_2$ coupling
when the two transitions meet.
The Ising-nematic transition (solid line) takes place at $T_{\sigma}$.
There can be either a first-order Ising-nematic and AFM(AFQ) transition at $T_{\sigma}=T_{AFM/AFQ}$,
or two separate transitions.}
\label{Fig:PhDTC}
\end{figure}

When
 interlayer
 bilinear-biquadratic couplings are taken into account, a phase with a pure AFQ
 order can be stabilized at finite temperature.
We can then discuss the evolution of the Ising-nematic transition as
a function of the $J_2/K_2$ ratio.
Consider the case when a dominating $J_2$
stabilizes a $(\pi,0)$ AFM order,
which is accompanied by the Ising-nematic order parameter $\sigma_1$.
For sufficiently large $K_2$, the AFQ order becomes the dominant order, and
the Ising-nematic order is predominantly given by $\sigma_2$.
The schematic evolution between the two limits is
illustrated in Fig.~\ref{Fig:PhDTC}.
We have illustrated the case with the quantum fluctuations having suppressed the weaker order.

We stress that,  such an evolution of the Ising-nematic transition
already occurs in the purely 2D model. Results from explicit calculations on the evolution of the
transition temperature $T_{\sigma}$
are shown in the Supplemental Material.\cite{SM}
In the case of the Ising-nematic transition associated with a $(\pi,0)$ AFM order,
the interlayer couplings give rise to a nonzero
$T_{AFM} \le T_{\sigma}$ (Refs.~\onlinecite{Fang08,Xu08,JDai09}).
Similarly, when the dominant order is a $(\pi,0)$ AFQ order,
such couplings lead to a nonzero $T_{AFQ} \le T_{\sigma}$.

{\it Implications for FeSe.~}General considerations suggest that
the cases of spin 1 or spin 2 are pertinent to the iron-based materials~\cite{SiAbrahams08}.
Judging from the measured total spin spectral
weight ~\cite{PCDai_review12},
the spin 1 case would
be closer to the iron pnictides while the spin 2 case would be more appropriate for the iron chalcogenides.

Accordingly,  it is natural to propose that the normal state of FeSe realizes the phase whose
ground state has the $(\pi,0)$ AFQ order accompanied by the Ising-nematic order.
In this picture, the structural transition at $T_s\sim90$ K corresponds to the concurrent
Ising-nematic and AFQ transition, as illustrated in Fig.~\ref{Fig:PhDTC}.
This picture explains why the structural phase transition is not accompanied by any  static AFM order.
At the same time, as soon as the AFQ order is developed, its Goldstone modes
will contribute towards low-energy dipolar magnetic fluctuations. This is consistent with the onset of
low-energy spin
fluctuations observed in the NMR measurements  \cite{Bohmer14,Baek14}.
It will clearly be important to explore such spin fluctuations using inelastic neutron
scattering measurements.
 A quantitative comparison between the measured and calculated spin excitation spectra would
 allow estimates of the coupling constants $J_n$ and $K_n$.
The Goldstone modes may also be probed by magnetoresistance, and unusual features in this property
have recently been reported \cite{Rossler14}.
Finally, the Ising-nematic order is linearly coupled not only to the structural anisotropy,
but also to the orbital order.
Similarly as for the iron pnictides~\cite{Yi11}, this would result in, for instance, the lifting of the
d$_{xz}$/d$_{yz}$ orbital degeneracy at the structural phase transition~\cite{Nakayama14,Shimojima14,Coldea14}.

The phase diagrams given in Fig.~\ref{Fig:Crossover} show that the AFQ region can be tuned
to an AFM region. The nature of the AFM phase depends on the bilinear couplings.
For a range of bilinear couplings, the nearby AFM phase has the ordering
wavevector $(\pi/2,\pi/2)$. This provides a means to connect the magnetism of FeSe and
FeTe~\cite{Bao,Li}, which is of considerable interest to the
on-going experimental efforts in studying the magnetism of the Se-doped FeTe series~\cite{Tranquada14}.
It also makes it natural to understand the development of magnetic order that seems to occur
when FeSe is placed under a pressure on the order of $1$ GPa ~\cite{Bendele10,Bendele12,Imai09}.
Finally, we note that our results will serve as the basis to shed new light on the nematic correlations in the superconducting state \cite{Song11,HungWu12,Chowdhury11}.

{\it Broader context.~} It is widely believed that understanding the magnetism
in the iron chalcogenide FeTe,
where the ordering wavvector $(\pi/2,\pi/2)$ has no connection with any Fermi-surface-nesting
features~\cite{Bao,Li}, requires a local-moment picture. The proposal advanced here not only
provides an understanding of the emerging puzzle on the magnetism in FeSe, but also
achieves a level of commonality in the underlying microscopic interactions across these iron
chalcogenides. Furthermore, the connection between the AFQ order
and the $(\pi,0)$ AFM order suggests that the local-moment physics, augmented by a coupling
to the coherent itinerant fermions near the Fermi energy, places the magnetism of a wide range
of iron-based superconductors in a unified framework. Since local-moment physics in bad metals
reflects a proximity to correlation-induced electron localization, this unified perspective
also signifies the importance of electron correlations \cite{SiAbrahams08,Yin,KSeo,Moreo,Lv,Yu2013}
 to the iron-based superconductors.

{\it Conclusions.~}To summarize, we have studied a generalized Heisenberg model with frustrated
bilinear-biquadratic
 interactions on a square lattice and find that the zero-temperature phase diagram
stabilizes an antiferroquadrupolar order. The anisotropic spin quadrupolar fluctuations
give rise to a finite-temperature Ising-nematic transition.
We propose that the structural phase transition in FeSe corresponds to this Ising-nematic transition
and is accompanied by an antiferroquadrupolar ordering.
We suggest that inelastic neutron scattering experiments be carried out to explore the proposed
Goldstone modes associated with the antiferroquadrupolar order.
Our results provide a natural
understanding for an emerging puzzle on FeSe.
More generally,
the extended phase diagrams advanced here considerably broaden the perspective on
the magnetism and electron correlations of the iron-based superconductors.

{\it Note added.~} Recently,
a study appeared that also emphasized the
local-moment-based magnetic physics for FeSe, but invoked a different mechanism
based on a possible paramagnetic Ising-nematic ground state caused by $J_1$-$J_2$
frustration~\cite{FWang15}. A distinction of the mechanism advanced here is that the AFQ order
yields Goldstone modes and therefore causes the onset of low-energy dipolar magnetic fluctuations.
In addition, results from inelastic neutron scattering experiments in FeSe have appeared~\cite{Rahn15,WangZhao15}, which
verify the $(\pi,0)$ magnetic excitations expected from our theoretical proposal.

We would like to acknowledge an early conversation with
C. Meingast, A. B\"ohmer and F. Hardy, which stimulated our interest in this problem,
and useful discussions with E. Abrahams, B. B\"{u}chner, A. Coldea, P. Dai, D.-H. Lee, and A. H. Nevidomskyy.
This work was supported in part by NSF Grant No.\ DMR-1309531, Robert A.\ Welch Foundation Grant No.\ C-1411
and the Alexander von Humboldt Foundation.
R.Y. was partially supported by National Science Foundation of China
Grant No. 11374361, and the Fundamental Research Funds for the
Central Universities and the Research Funds of Renmin University
of China. Both of us acknowledge the support provided in part by NSF Grant No. NSF PHY11-25915 at KITP, UCSB, for our participation in the Fall 2014
program on ``Magnetism, Bad Metals and Superconductivity: Iron Pnictides and Beyond."
Q.S.\  also acknowledges the hospitality of the the Karlsruhe Institute of Technology,
the Aspen Center for Physics (NSF Grant No.\ 1066293),
and the Institute of Physics of the Chinese Academy of Sciences.



\newpage

\setcounter{figure}{0}
\makeatletter
\renewcommand{\thefigure}{S\@arabic\c@figure}

\section{SUPPLEMENTARY MATERIAL --
Antiferroquadrupolar and Ising-nematic orders of a frustrated bilinear-biquadratic
Heisenberg model and implications for the magnetism of FeSe
}
\subsection{The ground-state spin configurations in the classical spin model}
Exactly at $T=0$, the classical $O(3)$ spins are always ordered. Therefore, the AFQ order is accompanied by magnetic dipolar orders.
Because the
$(\pi,0)$ AFQ order doubles the unit cell, the structure factor $\mathcal{S}_D(\mathbf{q})$ of the compatible magnetic dipolar order must show a two-$q$ structure as the consequence of Brillion zone folding, \emph{i.e.},
$\mathcal{S}_D(\mathbf{q})=\mathcal{S}_D(\mathbf{q}+\mathbf{q}_A)$
where $\mathbf{q}_A=(\pi,0)$.
The
ordering wavevector $\mathbf{q}$ depends on model parameters.

\begin{figure}[thb]
\includegraphics[width=80mm]{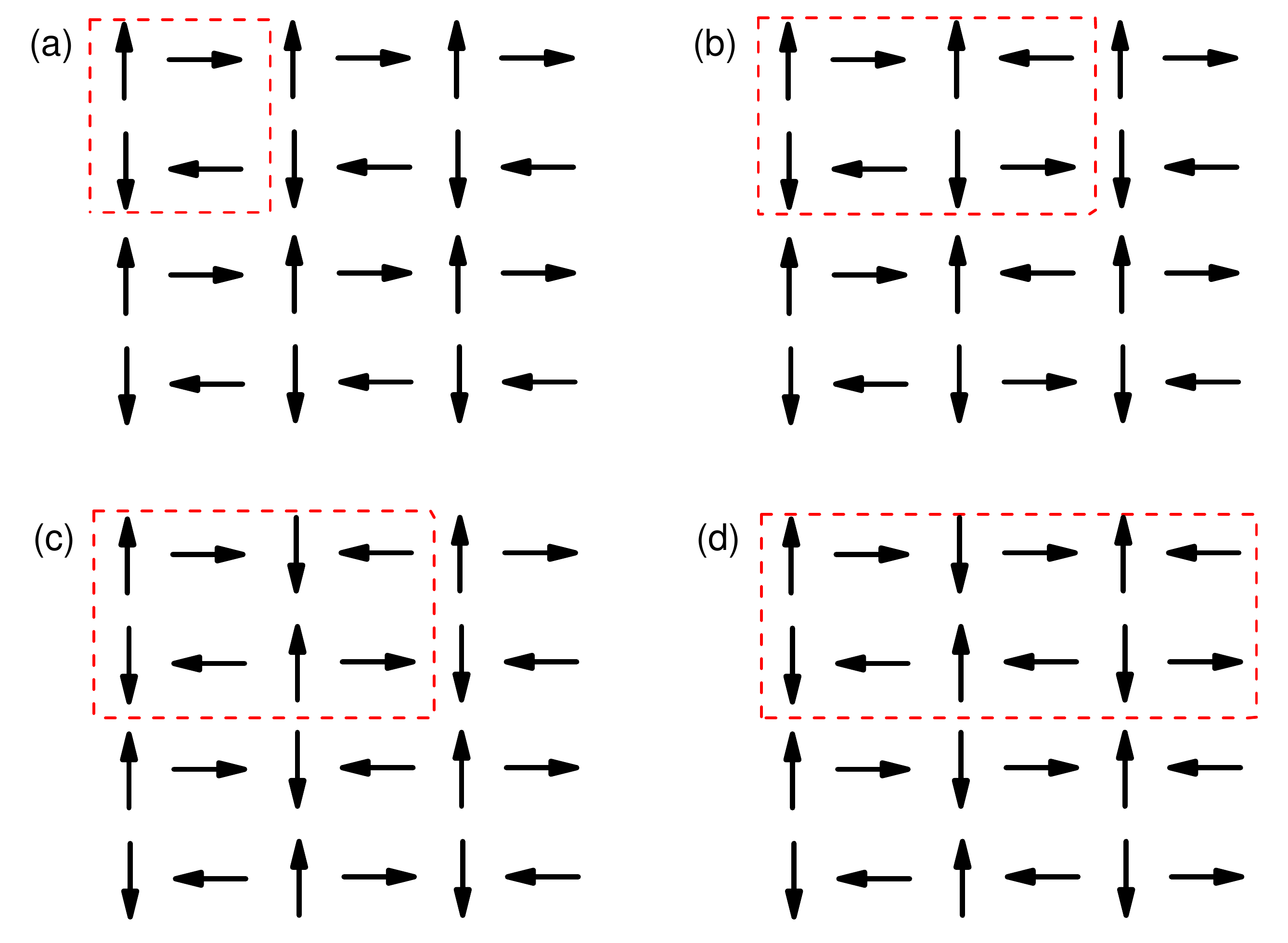}
\caption{Four among the infinitely degenerate ground-state spin patterns in the case of $J_3=0$ and $J_1=J_2$.
In each case, the
dashed box shows the magnetic unit cell.
The corresponding ordering wavevectors are as follows: $\mathbf{q}=(0,\pi)$ and $(\pi,\pi)$ in (a); $\mathbf{q}=(\pm\pi/2,\pi)$ in (b) and (c); $\mathbf{q}=(2\pi/3,\pi)$ and $(-\pi/3,\pi)$ in (d).}
\label{Fig:SpinConfig1}
\end{figure}

We find that in the $(\pi,0)/(0,\pi)$ AFQ ground state the spins are ordered at a wavevector $(q,\pi)/(\pi,q)$ with infinite degeneracies for $J_3=0$. Assuming a $(\pi,0)$ AFQ order, the spin variable at site $i$ is $\mathbf{S}_i = \xi(i_x)[\hat{e}_x\gamma(i_x) + \hat{e}_y(1-\gamma(i_x))]\gamma(i_y)$, where $i_x$ and $i_y$ are coordinates of site $i$, $\gamma(i_{x(y)}) = (1+(-1)^{i_{x(y)}})/2$, and $\xi(i_x)=\pm1$ is a random variable defined on each column of the lattice. The randomness in the real-space spin configuration leads to infinite number of degenerate ground-state spin patterns. Transforming to the momentum space, they correspond to ordering wavevector at $(q,\pi)$ (and $(q+\pi,\pi)$) with $q$ an arbitrary number. Some of the degenerate spin patterns are shown in Fig.~\ref{Fig:SpinConfig1}.

\begin{figure}[thb]
\includegraphics[width=80mm]{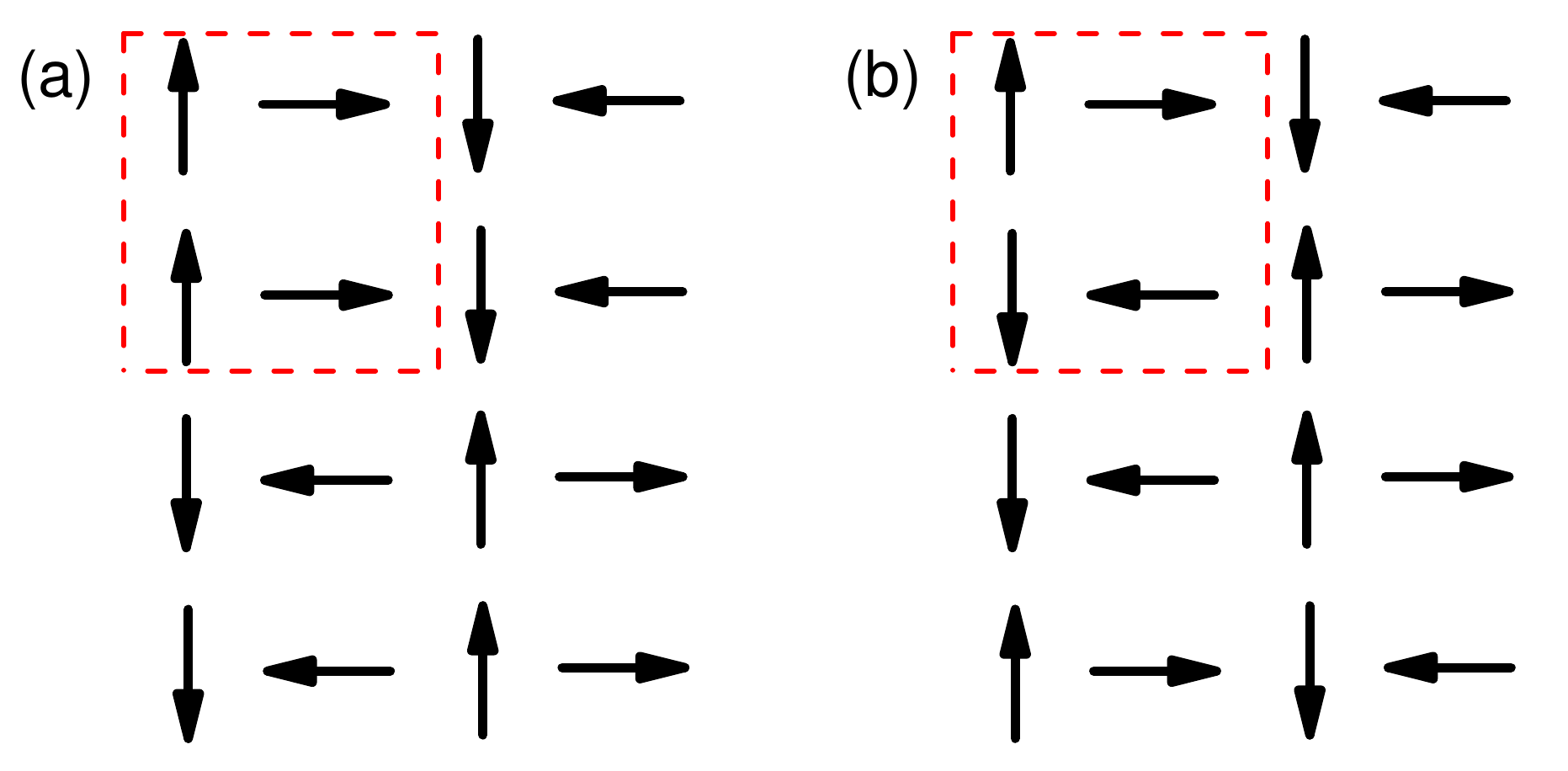}
\caption{Two out of the $16$ degenerate ground-state spin patterns in the case of $J_1=0$ and $J_2=J_3$. In each case, the spin pattern can be obtained by repeatedly aligning the spins in the dashed box in a staggered way.
In both cases, the
ordering wavevectors are $\mathbf{q}=(\pm \pi/2,\pm \pi/2)$ .}
\label{Fig:SpinConfig2}
\end{figure}

As for the case of $J_1=0$ and $J_2=J_3$, in the AFQ phase, we still find 16-fold degenerate ground-state spin patterns at ordering wavevectors $(\pm \pi/2,\pm \pi/2)$. Some of the spin patterns are shown in Fig.~\ref{Fig:SpinConfig2}.
In both cases, the large number of degenerate classical spin ground states helps to stabilize an AFQ order without a magnetic dipolar one when the quantum fluctuations are taken into account.

Another interesting observation is that at $T=0$, the dominant $(\pi,0)$ AFQ order of $Q^{x^2-y^2}$ coexists with a subleading FQ order of $Q^{r^2-3z^2}$. This can be checked using the ground-state spin configurations in Figs.~\ref{Fig:SpinConfig1} and \ref{Fig:SpinConfig2}, which gives $Q^{r^2-3z^2}_i=1/\sqrt{3}$ at each site and $Q^{x^2-y^2}=\pm1$.

\subsection{Spin excitations and Goldstone modes in the quantum $S=1$ model}
For the case of quantum spin $S=1$, the model defined in Eq. (1) of the main text can be studied by
an
SU(3) Schwinger boson
approach.\cite{SMLauchliMilaPenc06} At each site, let $|-1\rangle$, $|0\rangle$, and $|1\rangle$ be the three eigenstates of the spin operator $S^z$. We can define a time-reversal invariant basis of the SU(3) representation:
\begin{align}
 |x\rangle &= \frac{i}{\sqrt{2}} \left( |1\rangle -|-1\rangle\right), \nonumber \\
 |y\rangle &= \frac{1}{\sqrt{2}} \left( |1\rangle +|-1\rangle\right), \tag{S1}\\
 |z\rangle &= -i|0\rangle.\nonumber
\end{align}
Within this representation, we can then define three Schwinger bosons associated with the above three states, $b^\dagger_{\alpha}|\slashed{0}\rangle=|\alpha\rangle$, where $\alpha=x$, $y$, $z$, and $|\slashed{0}\rangle$ is
the
null state of the Schwinger bosons.
The three bosons satisfy a local constraint at each site:
\begin{equation}\label{Eq_constraint}
 \sum_\alpha b^\dagger_{i\alpha} b_{i\alpha} = 1. \tag{S2}
\end{equation}
The spin dipolar and quadrupolar operators can be written in terms of the Schwinger boson bilinears as
\begin{align}
  S^\alpha_i &= -i\epsilon_{\alpha\beta\gamma} (b^\dagger_{i\beta} b_{i\gamma} - b^\dagger_{i\gamma} b_{i\beta}), \tag{S3}\label{Eq:Sboson}\\\nonumber\\
  Q^{\alpha\beta}_i &= -(b^\dagger_{i\alpha} b_{i\beta} + b^\dagger_{i\beta} b_{i\alpha}),\nonumber\\
  Q^{x^2-y^2}_i &= -(b^\dagger_{ix}b_{ix}-b^\dagger_{iy}b_{iy}), \tag{S4}\label{Eq:Qboson}\\
  Q^{r^2-3z^2}_i &= \frac{1}{\sqrt{3}}(2b^\dagger_{iz}b_{iz}-b^\dagger_{ix}b_{ix}-b^\dagger_{iy}b_{iy}),\nonumber
\end{align}
where $\alpha$, $\beta$, and $\gamma$ run over $x$, $y$, and $z$, and  $\epsilon_{\alpha\beta\gamma}$ is the Levi-Civita symbol.
The Hamiltonian is then rewritten
as
\begin{align}\label{Eq:HamBoson}
 H &= \frac{1}{2}\sum_{i,\delta_n,\alpha,\beta} \left\{J_n b^\dagger_{i\alpha}b_{j\alpha}b^\dagger_{j\beta}b_{i\beta} + (K_n-J_n) b^\dagger_{i\alpha}b^\dagger_{j\alpha}b_{j\beta}b_{i\beta}\right\},\nonumber\\
 \tag{S5}
\end{align}
where
$j=i+\delta_n$, and $\delta_n$
(with $n=1$, $2$, $3$)
connects site $i$ and its $n$'s nearest neighbor sites.

We assume the following ground state at the mean level: $|\psi\rangle = \Pi_i |\psi_i\rangle = \Pi_i\sum_\alpha f_{i\alpha} b^\dagger_{i\alpha}|\slashed{0}\rangle$, where the coefficients satisfy $\sum_\alpha |f_{i\alpha}|^2=1$. The $(\pi,0)$
AFQ
order can be obtained by requiring condensation of $b_x$ and $b_y$ bosons at sites in odd and even columns, respectively. Correspondingly, the mean-field ground-state wave function at site $i$ is $|\psi_i\rangle = |x\rangle_i$ if the $x$ coordinate of site $i$ ($i_x$) is odd, and $|\psi_i\rangle = |y\rangle_i$ if $i_x$ is even. One could check that this wave function is indeed associated to an AFQ order at wavevector $\mathbf{q}_A=(\pi,0)$ since $\langle\psi_i|Q^{x^2-y^2}|\psi_i\rangle = e^{\mathbf{q}_A\cdot\mathbf{r}_i}=\pm1$.

We study the spin excitations in the AFQ phase by using a flavor-wave theory. We first perform a local rotation in the spin space,
\begin{align}
 \left(
 \begin{tabular}{c}
  $d_{ix}$\\
  $d_{iy}$\\
  $d_{iz}$
 \end{tabular}
 \right)
 &=
 \left(
 \begin{tabular}{ccc}
  $\cos \theta_i$ & $\sin \theta_i$ & $0$\\
  $-\sin \theta_i$ & $\cos \theta_i$ & $0$\\
  $0$ & $0$ & $1$
 \end{tabular}
 \right)
 \left(
 \begin{tabular}{c}
  $b_{ix}$\\
  $b_{iy}$\\
  $b_{iz}$
 \end{tabular}
 \right),\tag{S6}
\end{align}
such that in the rotated basis, only one flavor of bosons, $d_{ix}$, condenses. In the $(\pi,0)$ AFQ phase, this corresponds to taking $\theta_i=0$ if $i_x$ is odd and $\theta_i=\pi/2$ if $i_x$ is even.
Using the constraint in this rotated basis, $\sum_\alpha d^\dagger_{i\alpha} d_{i\alpha}=1$, we obtain
\begin{align}\label{Eq:FWA}
d_{ix} &\rightarrow \sqrt{1 - d^\dagger_{iy}d_{iy} - d^\dagger_{iz}d_{iz}}\nonumber\\
 &\approx 1-\frac{1}{2}(d^\dagger_{iy}d_{iy} + d^\dagger_{iz}d_{iz}).\tag{S7}
\end{align}

Using Eq.~\eqref{Eq:FWA}, we can expand the Hamiltonian in Eq.~\eqref{Eq:HamBoson} in terms of the magnon operators $d_{iy}$, $d_{iz}$, and their Hermitian conjugates.
We then truncate the expanded Hamiltonian to keep up to the quadratic terms of $d_{iy}$, $d_{iz}$, and their Hermitian conjugates. Given that the ground state is the AFQ state, the linear terms in $d_{iy}$, $d_{iz}$ automatically cancel out, and we
 arrive at, up to a constant energy, a quadratic Hamiltonian. After Fourier
transforming
 it to the momentum space, this quadratic Hamiltonian reads,
\begin{align}\label{Eq:Ham2b}
 H \approx \sum_{k,\alpha=y,z} & A_{k\alpha} (d^\dagger_{k\alpha}d_{k\alpha} + d_{-k\alpha}d^\dagger_{-k\alpha})\nonumber\\
  +& B_{k\alpha} (d^\dagger_{k\alpha}d^\dagger_{-k\alpha} + d_{k\alpha}d_{-k\alpha}). \tag{S8}
\end{align}
Here $k$ runs over the unfolded Brilluion zone (BZ), and
\begin{align}
 A_{ky} &= J_1(\cos k_x+ \cos k_y) - K_1 \cos k_x \nonumber\\
 & - 2(K_2-J_2) \cos k_x \cos k_y + J_3 (\cos 2k_x+ \cos 2k_y) \nonumber\\
 & + 2K_2 -2K_3,\tag{S9}\\ \nonumber\\
 B_{ky} &= K_1\cos k_y - J_1(\cos k_x+\cos k_y) - 2J_2 \cos k_x \cos k_y \nonumber\\
 &+ (K_3-J_3) (\cos 2k_x+ \cos 2k_y),\tag{S10}\\ \nonumber\\
 A_{kz} &= J_1 \cos k_y +J_3(\cos 2k_x+ \cos 2k_y) -K_1 -2K_3, \tag{S11}\label{Eq:Akz}\\ \nonumber\\
 B_{kz} &= (K_1-J_1)\cos k_y + (K_3-J_3) (\cos 2k_x+ \cos 2k_y). \tag{S12}\label{Eq:Bkz}
\end{align}
The Hamiltonian in Eq.~\eqref{Eq:Ham2b} can be diagonalized via a Bogoliubov transformation
\begin{align}
 d_{ky(z)} &= u_{ky(z)} \gamma_{ky(z)} - v_{ky(z)} \gamma_{ky(z)}^\dagger, \tag{S13}
\end{align}
and
\begin{align}
 H &\approx \sum_{k,\alpha=y,z}\epsilon_{k\alpha} (\gamma^\dagger_{k\alpha} \gamma_{k\alpha} + \gamma_{-k\alpha} \gamma_{-k\alpha}^\dagger), \tag{S14}
\end{align}
where the magnon dispersion
\begin{align}\label{Eq:disp}
 \epsilon_{k\alpha} &= \sqrt{A^2_{k\alpha}-B^2_{k\alpha}}, \tag{S15}
\end{align}
and
\begin{align}
 u_{k\alpha} &= \frac{1}{2}\left( \frac{A_{k\alpha}}{\sqrt{A^2_{k\alpha}-B^2_{k\alpha}}} + 1 \right), \nonumber\\
 v_{k\alpha} &= \frac{1}{2}\left( \frac{A_{k\alpha}}{\sqrt{A^2_{k\alpha}-B^2_{k\alpha}}} - 1 \right).\tag{S16}
\end{align}

The parameter regime where the $(\pi,0)$ AFQ phase is stable is obtained by requiring $A_{k\alpha}\geqslant0$ and $A^2_{k\alpha}-B^2_{k\alpha}\geqslant0$ at every $k$ in the
entire BZ and for both $\alpha=y,z$, and can be determined numerically.

The dynamical spin dipolar and quadrupolar structure factors $S^{\mu\mu}_Q(\mathbf{q},\omega)$ and $S^{\alpha\alpha}_D(\mathbf{q},\omega)$ can be calculated within the diagonalized
representation. In gerneral,
\begin{align}
 S^{\mu\mu}_Q(\mathbf{q},\omega) & = \sum_m |\langle g|Q^\mu_{\mathbf{q}}|m\rangle|^2 \delta(\omega-(E_m-E_g)),\tag{S17}\\
 S^{\alpha\alpha}_D(\mathbf{q},\omega) & = \sum_m |\langle g|S^\alpha_{\mathbf{q}}|m\rangle|^2 \delta(\omega-(E_m-E_g)).\tag{S18}
\end{align}
Here, $|g\rangle$ and $|m\rangle$ refer to the ground state (with eigenenergy $E_g$) and the $m$'s excited state (with eigenenergy $E_m$) in the flavor-wave theory.
\begin{align}
 Q^\mu_{\mathbf{q}} = \frac{1}{\sqrt{N}} \sum_i e^{i\mathbf{q}\cdot\mathbf{r}_i} Q^\mu_i,\tag{S19}\\
 S^\alpha_{\mathbf{q}} = \frac{1}{\sqrt{N}} \sum_i e^{i\mathbf{q}\cdot\mathbf{r}_i} S^\alpha_i,\tag{S20}
\end{align}
where $N$ is the number of spins of the system, and $Q^\mu_i$ and $S^\alpha_i$ are expressed in terms of Schwinger bosons using Eqs.~\eqref{Eq:Sboson} and ~\eqref{Eq:Qboson}. For example, for $\mu=x^2-y^2$,
\begin{align}
 Q^{x^2-y^2}_{\mathbf{q}} &= \sqrt{N} \delta_{\mathbf{q},\mathbf{q}_A} - \frac{2}{\sqrt{N}} \sum_k d^\dagger_{ky}d_{k+\mathbf{q}+\mathbf{q}_A,y} \nonumber\\
 &- \frac{1}{\sqrt{N}} \sum_k d^\dagger_{kz}d_{k+\mathbf{q}+\mathbf{q}_A,z}.\tag{S21}
\end{align}
This leads to $S^{\mu\mu}_Q(\mathbf{q},\omega) = N \delta_{\mathbf{q},\mathbf{q}_A} \delta(\omega)$ up to the one-magnon contribution, which confirms the AFQ order at $\mathbf{q}_A=(\pi,0)$.

The one-magnon contribution to the spin dipolar correlation function is also non-zero. We find that the transverse dynamical structure factor
\begin{align}
 S^{xx}_D(\mathbf{q},\omega) = S^{yy}_D(\mathbf{q},\omega) &= \frac{1}{2} \sqrt{\frac{A_{\mathbf{q}z}+B_{\mathbf{q}z}}{A_{\mathbf{q}z}-B_{\mathbf{q}z}}} \delta(\omega-\epsilon_{\mathbf{q}z})\tag{S22}
\end{align}
From Eqs.~\eqref{Eq:Akz},\eqref{Eq:Bkz}, and \eqref{Eq:disp}, we find a Goldstone mode near $\mathbf{q}_A=(\pi,0)$. For $\mathbf{q}=\mathbf{q}_A+(dq_x,dq_y)$,
\begin{align}
 \epsilon_{\mathbf{q}z} &\approx \sqrt{v_x^2dq_x^2+v_y^2dq_y^2},\tag{S23}
\end{align}
with anisotropic velocities
\begin{align}
 v_x&=2\sqrt{K_3\eta}, \nonumber\\
 v_y&=\sqrt{(K_1+4K_3)\eta}, \tag{S24}
\end{align}
where $\eta=K_1+2K_3-J_1-2J_3$.
The Goldstone mode near $\mathbf{q}_A$ is also seen in the dynamical spin dipolar structure factor:
\begin{align}
 S^{xx}_D(\mathbf{q}\to\mathbf{q}_A, \omega)\approx \frac{\omega}{2\eta} \delta(\omega-\epsilon_{\mathbf{q}z}). \tag{S25}
\end{align}
Note that the static dipolar structure factor $S^{xx}_D(\mathbf{q}_A)=0$, because of the absence of long-range magnetic dipolar order in the AFQ phase. But the Goldstone modes associated with the broken spin rotational symmetry can be observed from the dynamical spin dipolar structure factor. A representative plot of $S^{xx}_D(\mathbf{q}, \omega)$ showing the Goldstone modes is displayed in Fig.~4 of the main text.

\subsection{Quantum fluctuations and coupling to itinerant fermions.~}
The field theory that describes the two coupled order parameters $\mathbf{Q}_A$
and $\mathbf{Q}_B$ will be similar to that of the $(\pi,0)$
AFM order of the $J_1-J_2$
Heisenberg
model.
The effect of coupling to the itinerant fermions can be treated as in
Ref.~\cite{SMJDai09}
within
an effective Ginzburg-Landau action:
\begin{align}
 {\cal S}({\bf Q}_A,{\bf Q}_B) &=
 \int d\{\bf q\}
  \int d\{\omega\} [ {\cal S}_{2}
  + {\cal S}_{4}
  + \dots], \nonumber \\
{\cal S}_2({\bf q},\omega) &=
r(\omega,\mathbf{q})
[|{\bf Q}_A({\bf q},\omega)|^2 + |{\bf Q}_B({\bf q},\omega)|^2] \nonumber \\
& +  v(q_x^2-q_y^2)[{\bf Q}_A({\bf q},\omega)\cdot {\bf Q}_B({-\bf q},-\omega)], \nonumber \\
{\cal S}_4 (\{{\bf q}\},\{\omega\}) &= u(|{\bf Q}_A|^4 + |{\bf Q}_B|^4)
+ u'|{\bf Q}_A|^2\; |{\bf Q}_B|\,^2
\nonumber\\
& + {\tilde u}|{\bf Q}_A\cdot {\bf Q}_B|^2 . \tag{S26}
\label{Eq:GL}
\end{align}
where $r(\omega,\mathbf{q}) =r_0+wA+\omega^2+\gamma|\omega| + c q^2 $,
$r_0<0$ and $A>0$ are constants,
$w$ is the coherent quasiparticle spectral weight of
itinerant electrons, and $\gamma$
is a Landau-damping coefficient.
Note that ${\tilde u}<0$, and
$\{{\bf q}\},\{\omega\}$
mark
the set of four momenta and four frequencies that enter $S_4$;
the
momentum and frequency integrals
 are understood to each contain a delta function that fixes the sum of the momenta and the sum of the frequencies at zero.
A similar form also exists for the AFM orders, $\mathbf{M}_A$ and $\mathbf{M}_B$
~\cite{SMJDai09}.
The shift of $r$ by $w$ and the damping may lead to the loss of magnetic order in the system~\cite{SMJDai09},
and stabilize a pure AFQ order.
In the absence of the AFQ order, the Ising-nematic order will be concurrently suppressed \cite{SMJWu14}.
However, when the dominant order is AFQ, one can readily reach the regime where
the quantum fluctuations eliminate the weaker AFM order while retaining the stronger AFQ order and the associated Ising-nematic order.

\begin{figure}[t!]
\includegraphics[width=80mm]{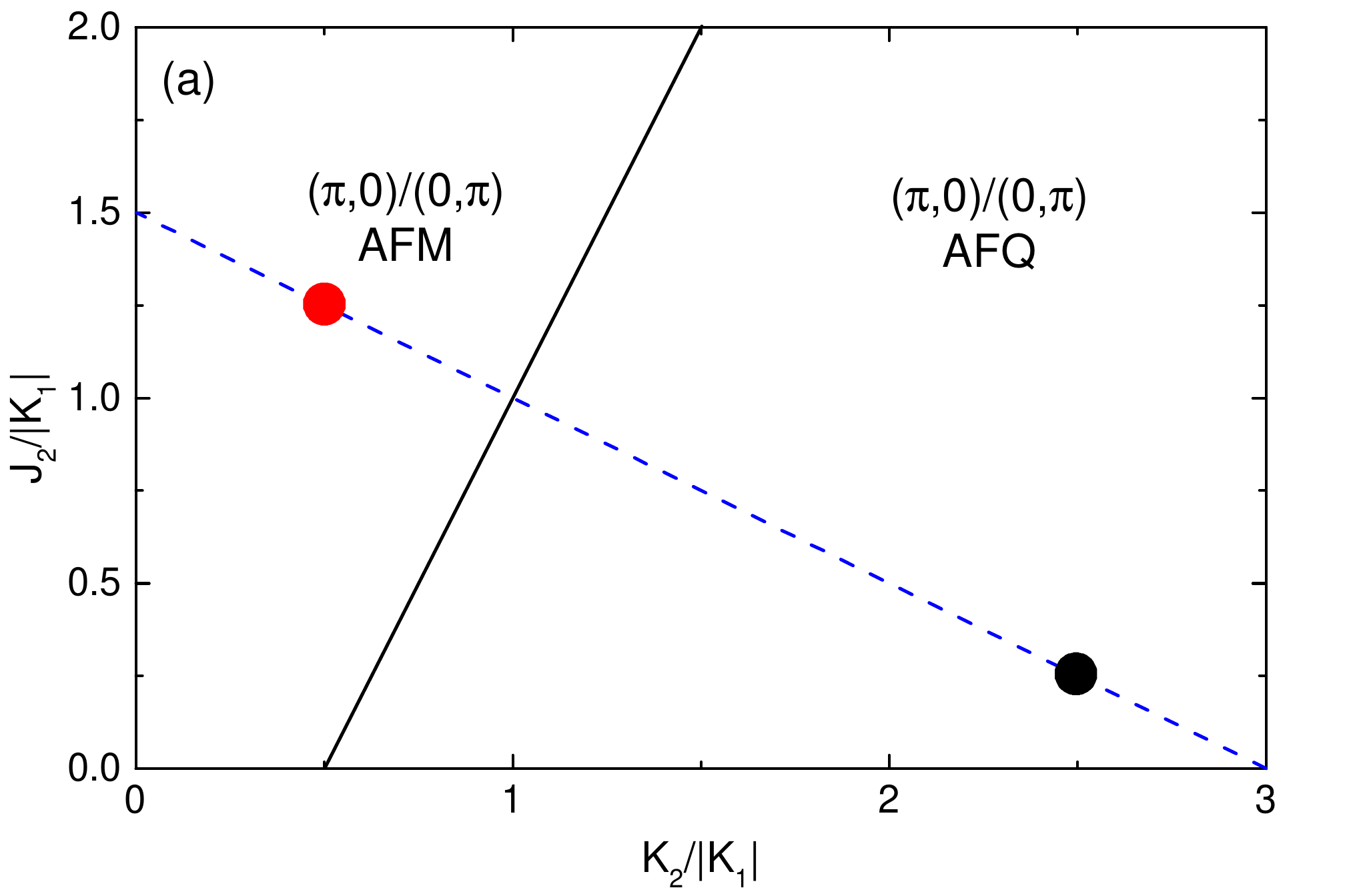}
\includegraphics[width=80mm]{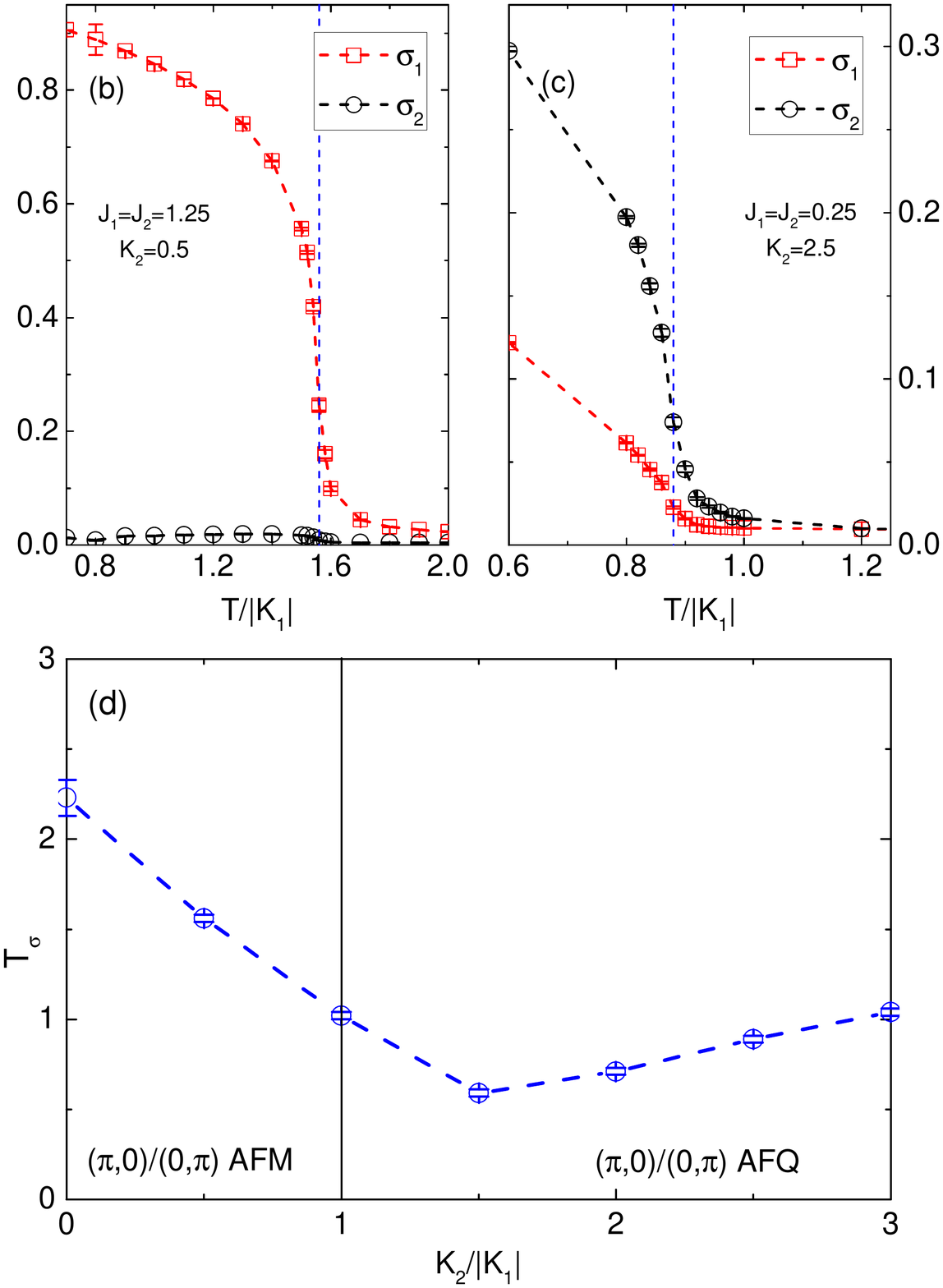}
\caption{
(a): Mean-field ground-state phase diagram of the classical
bilinear-biquadratic Heisenberg model at $J_3=K_3=0$ and $J_1=J_2$.
The blue dashed line shows the trajectory by tuning $J_2/K_2$, and the red and black dots show two representative points along this trajectory.
(b) and (c): Temperature dependence of
the Ising-nematic order parameters $\sigma_1$ and $\sigma_2$ at the two representative points of the phase diagram in panel (a).
There is a change of the dominant Ising-nematic order parameter from $\sigma_1$ to $\sigma_2$ when the ground state varies from the AFM to the AFQ phase.
The blue dashed line marks the
transition
temperature, $T_\sigma$, of the Ising-nematic transition. Data shown are based on Monte Carlo simulations on a $40\times 40$ lattice.
(d): Evolution of $T_\sigma$ with varying $J_2/K_2$ along the trajectory
shown in panel (a). A minimum of $T_\sigma$ is located near the phase boundary (black solid line) between the AFM and the AFQ phase, which is consistent with the
proposal made in the main text, as illustrated
in Fig. 5 of the main text.
}
\label{Fig:IsingEvol}
\end{figure}

\subsection{The evolution of the
Ising-nematic
order parameter and transition temperature
as a function of $J_2/K_2$}
Here we study how the dominant Ising-nematic order parameter and
the associated transition temperature,
$T_\sigma$, change with varying the $J_2/K_2$ ratio in the 2D classical
bilinear-biquadratic Heisenberg
model. The calculations are done via classical Monte Carlo simulations using Metropolis sampling with up to $80\times80$ lattices and $10^5$ Monte Carlo steps.

We tune the
 $J_2/K_2$ ratio by going along the blue dashed trajectory in the phase diagram shown in Fig.~\ref{Fig:IsingEvol}(a). As discussed in the main text, with increasing $K_2$, the ground state of the system changes from the $(\pi,0)/(0,\pi)$ AFM to the $(\pi,0)/(0,\pi)$ AFQ state. We have tracked the evolution of the Ising-nematic order parameters $\sigma_1$ and $\sigma_2$, and find that the change of
 the
 ground state is reflected in the variation of the Ising-nematic order parameters: the dominant Ising-nematic order parameter changes from $\sigma_1$ in the AFM phase to $\sigma_2$ in the AFQ phase, as clearly shown in Fig.~\ref{Fig:IsingEvol}(b) and (c). We have also determined the
 transition
 temperature $T_\sigma$ from our numerical results. $T_\sigma$ first decreases then increases with increasing $K_2$ along the trajectory, showing a
remarkable
 minimum near the phase boundary between the AFM and AFQ phases (Fig.~\ref{Fig:IsingEvol}(d)).
 This minimum indicates enhanced fluctuations around the Ising-nematic order due to the competition between the AFM and AFQ orders.
 Our
 results on the evolution of the dominant Ising-nematic order parameter and $T_\sigma$ with $J_2/K_2$ are fully consistent with the general
picture
 proposed in Fig. 5 of the main
 test.
When an interlayer coupling is turned on, we
 expect similar results
 for the Ising-nematic order,
 because
 the
 change of the dominant Ising-nematic order parameter and the evolution of $T_\sigma$ reflect the competition between the underlying AFM and AFQ ground states.
 In this case, in the dominating AFM regime, it is well known that
 a nonzero transition temperature develops for the AFM order. Similar reasoning applies to the dominating AFQ regime.

\end{document}